\let\myover=\over      
\documentclass[10pt,a4paper]{article}
\usepackage{amssymb,amsmath,epsfig,graphics}
\def\e{{\rm e}}

\def\d{\partial}
\def\l{\left(}
\def\r{\right)}
\def\la{\langle }
\def\ra{\rangle }
\newcommand{\be}{\begin{equation}}
\newcommand{\ee}{\end{equation}}

\renewcommand{\ln}{\mathop{\rm ln}\nolimits}

\newcommand{\bg}{\begin{gather}}
\newcommand{\eg}{\end{gather}}

\def\half{\frac{1}{2}}

\newcommand{\goe}{\gtrsim}
\newcommand{\loe}{\lesssim}
\begin{document}
\let\over=\myover  
\def\half{{1 \over 2}} 
\title{Kaon Physics with Light Sgoldstinos and Parity Conservation}
\author{ D.~S.~Gorbunov$^a$\thanks{{\bf e-mail}: gorby@ms2.inr.ac.ru}, 
V.~A.~Rubakov$^{a,b}$\thanks{{\bf e-mail}: rubakov@ms2.inr.ac.ru}\\
{\small{\em
$^a$~Institute for Nuclear Research of the Russian Academy of Sciences, }}\\
{\small{\em
60th October Anniversary prospect 7a, Moscow 117312, Russia}} \\
{\small{\em
$^{b}$~Max-Plank-Institut f\"ur Physik, Werner-Heisenberg-Institut,}}\\
{\small{\em
F\"ohringer Ring 6, 80805, Munich, Germany
}}}
\date{}
\maketitle
\begin{abstract} 
  \small 
Superpartners of goldstino --- scalar and pseudoscalar sgoldstinos ---
interact weakly with ordinary particles. One or both of them may be
light. We consider a class of supersymmetric extensions of the
Standard Model in which interactions of sgoldstinos with
quarks and gluons conserve parity but do not conserve
quark flavor. If the pseudoscalar sgoldstino $P$ is light,
$m_P <(m_K - 2 m_{\pi})$, and the scalar sgoldstino is heavier,
$m_S > (m_K - m_{\pi})$, an interesting place for experimental
searches is the poorly explored area of three-body decays of kaons,
$K^{0}_{S,L} \to \pi^+ \pi^- P$, $K^{0}_{S,L} \to \pi^0 \pi^0 P$ and
 $K^{+} \to \pi^+ \pi^0 P$, with $P$ subsequently decaying into
$\gamma \gamma$, possibly  $e^+ e^-$, or flying away from the
detector. We evaluate the constraints on the flavor-violating coupling
of sgoldstino to quarks which are imposed by
 $K_L^0 - K_S^0$ mass difference and
CP-violation in neutral kaon system, and find that these 
constraints allow 
for fairly large $\mbox{Br}(K \to \pi \pi P)$. 
Depending on the phase of sgoldstino-quark coupling, most sensitive to
light pseudoscalar sgoldstino are searches either for decays
$K_L^0\to\pi\pi P$ or $K^+\to\pi^+\pi^0P$ and $K_S^0\to\pi\pi
P$. Generally speaking, there are no bounds on ${\rm Br}(K^0_L \to \pi \pi
P)$. For most values of the phase, branching ratio of
$K^+\to\pi^+\pi^0P$ is about three orders of magnitude smaller than 
${\rm Br}(K_L^0\to\pi\pi P)$ and the branching ratios of $K_S^0\to\pi\pi P$
are very small. However, for a certain phase the situation is
opposite. We find that the most
interesting ranges of branching ratios start at 
\[
\mbox{Br}(K^0_L \to \pi \pi P) \sim 10^{-3}\, , ~~~~
\mbox{Br}(K^+ \to \pi^+ \pi^0 P) \sim 10^{-4}
\, , ~~~~
\mbox{Br}(K_S^0 \to \pi \pi P) \sim 10^{-3}\, .
\]
These searches for light pseudoscalar sgoldstino 
would be sensitive to the supersymmetry breaking scale $\sqrt{F}$
in the 100 TeV range and above, provided MSSM flavor violating
parameters are close to their experimental bounds. 
We also briefly discuss the cases of
light scalar sgoldstino and relatively heavy sgoldstinos.  
\end{abstract}

~~~1. In 
supersymmetric models of particle physics, spontaneous supersymmetry 
breaking results in the appearance of a Goldstone fermion --- goldstino
--- which becomes the longitudinal component of gravitino. There should 
exist also superpartners of goldstino, pseudoscalar $P$ and scalar
$S$, both neutral under all gauge interactions. The masses of $P$ and $S$
are in general different; their values are model-dependent and may well 
be lower than a few GeV or even a few MeV. These bosons --- sgoldstinos ---
are indeed light in various versions of both gravity mediated 
theories~\cite{ellis,no-scale} and gauge mediated models (see,
e.g., Ref.~\cite{gmm} and references therein). It is certainly of 
interest to search for sgoldstinos at 
colliders~\cite{dicus,0001025,Abreu:2000ij} and in rare 
decays~\cite{Brignole:2000wd,low}.

Interactions of sgoldstinos with ordinary quarks, leptons and gauge 
bosons are suppressed by the scale, traditionally denoted by
$\sqrt{F}$, at which supersymmetry is broken in the underlying
theory. On the one hand, this means that sgoldstinos are naturally
weakly coupled to ordinary particles. On the other hand, sgoldstinos,
in similarity to gravitinos, are potential sources of information 
about this fundamental scale, which otherwise enters low energy 
physics indirectly, through soft supersymmetry breaking masses and 
couplings of ordinary  particles and their superpartners.

Below the electroweak scale, interactions of sgoldstinos with 
quarks and gluons may or may not conserve parity.
If parity is not conserved, there is no real distinction between
the pseudoscalar sgoldstino $P$ and scalar sgoldstino $S$ insofar
as
their couplings to hadrons are concerned. If parity is conserved,
the situation is different: we will see that the low energy 
phenomenology of the light pseudoscalar sgoldstino  is not completely 
standard.
In this paper we 
will mostly consider the case of parity-conserving sgoldstino
interactions, and comment on the opposite case in appropriate places.

Parity conservation in sgoldstino interactions with quarks and gluons
(as well as with leptons and photons) may not be accidental.
As an example, it is natural in theories with 
spontaneously broken left-right symmetry, as we discuss in Appendix 1.
We note in this regard that left-right symmetric extensions of MSSM
(for a review see, e.g., Ref.~\cite{Mohrev}) not only are aesthetically
appealing but also provide a solution~\cite{strongCP} to the strong 
CP-problem, which is a viable alternative to the Peccei--Quinn mechanism.
It is likely that sgoldstino interactions will conserve parity
in supersymmetric versions of other models 
(see, e.g., Ref.~\cite{otherstronCP}) designed to solve
the strong CP-problem
without introducing light axion\footnote{More realistically,
loop effects induce non-zero  parity-violating terms
in the low energy sgoldstino-quark Lagrangian even if such 
terms are absent at the tree level. In the context of
left-right models, we find in Appendix 1 that these loop 
contributions are small.}.

Parity-conserving  low energy
interactions of pseudoscalar sgoldstino
$P$ with quarks are written\footnote{If parity is not conserved in
sgoldstino-quark interactions, the particle $P$ 
couples to both
pseudoscalar and scalar densities, $\bar{q}_i i\gamma^5 q_j$
and $\bar{q}_i q_j$. If the scalar coupling is considerable, 
low energy phenomenology of $P$ is similar to that of 
$S$. The latter will be discussed towards the end of this paper.}
 as 
follows~\cite{bhat,9904367,0001025},
\begin{equation}
    {\cal L}_{P,q} =   
-P \cdot (h_{ij}^{(D)} \cdot \bar{d}_i\, i \gamma^5 d_j 
+ h_{ij}^{(U)} \cdot \bar{u}_i\, i\gamma^5 u_j)\;,
\label{3*}
\end{equation}
where
\begin{equation}
   d_i = (d,s,b)\, ,\,\,\,\, u_i = (u,c,t)\;.
\nonumber
\end{equation}
In general, 
the coupling constants $h_{ij}^{(D,U)}$ 
receive contributions from various terms in the Lagrangian of an
underlying theory. In particular, there are always contributions 
proportional to the
left-right soft terms in the matrix of squared masses of
squarks,\footnote{One of the conditions ensuring parity conservation 
in sgoldstino-fermion interactions is that the left-right soft
mass matrices are Hermitian, 
$[\tilde{m}_{D}^{(LR)2}]^{\dagger}=\tilde{m}_{D}^{(LR)2}$,
and similarly for $\tilde{m}_{U}^{(LR)2}$ and $\tilde{m}_{L}^{(LR)2}$,
where $\tilde{m}_{L}^{(LR)2}$ refers to sleptons.}
\begin{equation}
    h_{ij}^{(D)} = {1\over \sqrt{2}}
\frac{\tilde{m}_{D,ij}^{(LR)2}}{F}\;,
\label{hDD}
\end{equation}
\begin{equation}
    h_{ij}^{(U)} = {1\over \sqrt{2}}
\frac{\tilde{m}_{U,ij}^{(LR)2}}{F}\;.
\label{hUU}
\end{equation}
We will use Eqs.~(\ref{hDD}) and (\ref{hUU}) later on to estimate the
sensitivity of low energy experiments to the scale of supersymmetry 
breaking, $\sqrt{F}$.

The low energy interactions of  scalar sgoldstino $S$ are governed by 
the same coupling constants (again assuming parity conservation),
\begin{equation}
    {\cal L}_{S,q} = -S\cdot (h_{ij}^{(D)} \cdot \bar{d}_i d_j 
+ h_{ij}^{(U)} \cdot \bar{u}_i  u_j)\;.
\label{3**}
\end{equation}

\begin{figure}[htb]
\begin{picture}(0,0)%
\epsfig{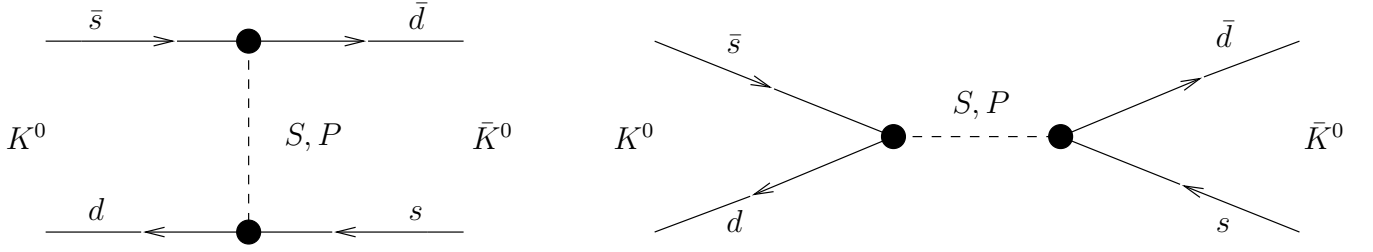}%
\end{picture}%
\setlength{\unitlength}{3947sp}%
\begingroup\makeatletter\ifx\SetFigFont\undefined%
\gdef\SetFigFont#1#2#3#4#5{%
  \reset@font\fontsize{#1}{#2pt}%
  \fontfamily{#3}\fontseries{#4}\fontshape{#5}%
  \selectfont}%
\fi\endgroup%
\begin{picture}(8550,1466)(151,-1044)
\put(8501,-436){\makebox(0,0)[lb]{\smash{\SetFigFont{12}{14.4}{\rmdefault}{\mddefault}{\updefault}$\bar{K}^0$}}}
\put(7951,-961){\makebox(0,0)[lb]{\smash{\SetFigFont{12}{14.4}{\rmdefault}{\mddefault}{\updefault}$s$}}}
\put(7951,199){\makebox(0,0)[lb]{\smash{\SetFigFont{12}{14.4}{\rmdefault}{\mddefault}{\updefault}$\bar{d}$}}}
\put(4166,-436){\makebox(0,0)[lb]{\smash{\SetFigFont{12}{14.4}{\rmdefault}{\mddefault}{\updefault}$K^0$}}}
\put(6296,-211){\makebox(0,0)[lb]{\smash{\SetFigFont{12}{14.4}{\rmdefault}{\mddefault}{\updefault}$S,P$}}}
\put(4876,164){\makebox(0,0)[lb]{\smash{\SetFigFont{12}{14.4}{\rmdefault}{\mddefault}{\updefault}$\bar{s}$}}}
\put(4876,-961){\makebox(0,0)[lb]{\smash{\SetFigFont{12}{14.4}{\rmdefault}{\mddefault}{\updefault}$d$}}}
\put(2101,-416){\makebox(0,0)[lb]{\smash{\SetFigFont{12}{14.4}{\rmdefault}{\mddefault}{\updefault}$S,P$}}}
\put(351,-436){\makebox(0,0)[lb]{\smash{\SetFigFont{12}{14.4}{\rmdefault}{\mddefault}{\updefault}$K^0$}}}
\put(3276,-436){\makebox(0,0)[lb]{\smash{\SetFigFont{12}{14.4}{\rmdefault}{\mddefault}{\updefault}$\bar{K}^0$}}}
\put(866,314){\makebox(0,0)[lb]{\smash{\SetFigFont{12}{14.4}{\rmdefault}{\mddefault}{\updefault}$\bar{s}$}}}
\put(866,-886){\makebox(0,0)[lb]{\smash{\SetFigFont{12}{14.4}{\rmdefault}{\mddefault}{\updefault}$d$}}}
\put(2881,314){\makebox(0,0)[lb]{\smash{\SetFigFont{12}{14.4}{\rmdefault}{\mddefault}{\updefault}$\bar{d}$}}}
\put(2881,-886){\makebox(0,0)[lb]{\smash{\SetFigFont{12}{14.4}{\rmdefault}{\mddefault}{\updefault}$s$}}}
\end{picture}
\caption{Sgoldstino contribution to $K_L^0-K_S^0$ mass difference
and CP-violation in the system of neutral kaons.}
\label{k-k-mixing}
\end{figure}

It has been pointed out in Refs.~\cite{Brignole:2000wd,low}
that
sgoldstino 
interactions generically 
violate quark flavor and CP.
Flavor changing processes and CP-violation
occur due to off-diagonal elements in the (Hermitian)
matrices of couplings $h_{ij}^{(D,U)}$; these off-diagonal
elements are generally complex.
In this note we will be primarily interested in kaon
physics, in which case the relevant flavor-violating terms in the
low energy Lagrangian are
\begin{equation}
   {\cal L}_{P,q} = -P\cdot(h_{12}^{(D)} \cdot \bar{d}\, i\gamma^5 s
     + \mbox{h.c.})\;,
\label{4+}
\end{equation}
\begin{equation}
   {\cal L}_{S,q} = -S\cdot(h_{12}^{(D)} \cdot \bar{d}  s
     + \mbox{h.c.})\;.
\nonumber
\end{equation}
These terms induce two major effects in kaon physics. First,
sgoldstino exchange contributes to $K_L^0 - K_S^0$ mass difference
and CP-violation in the system of neutral kaons, as shown 
in fig.~\ref{k-k-mixing}. 
Second, if sgoldstinos are sufficiently light, they can be produced 
in kaon decays. Let us discuss these two effects in turn.

~~~2. We begin with the pseudoscalar sgoldstino $P$. If it is light,
$m_P \loe m_K$, its contribution to $K_L^0 - K_S^0$ mass difference 
is readily calculated in chiral theory
\begin{equation}
  \Delta m_K \equiv m_{K_L^0} - m_{K_S^0} 
=[(\mbox{Re}h_{12}^{(D)})^2 - (\mbox{Im}h_{12}^{(D)})^2]
\frac{B_0^2 f_{K}^2}{m_K(m_K^2 - m_P^2)}\;,
\nonumber
\end{equation}
where $f_K = 160$~MeV and
the constant $B_0$ is related to quark condensate, 
$\la 0|\bar{q}q|0\ra=-\half B_0f_\pi^2$, $f_\pi=130$~MeV, 
that is $B_0=M_{K}^2/(m_d+m_s)=1.9$~GeV. 
Neglecting the mass of 
$P$ and requiring that sgoldstino contribution does
not exceed the actual value of
$\Delta m_K$, we obtain in the case of light pseudoscalar 
sgoldstino
\begin{equation}
|(\mbox{Re}~h_{12}^{(D)})^2 - (\mbox{Im}~h_{12}^{(D)})^2|
< 5\cdot 10^{-15}\;.
\label{full}
\end{equation}
In what follows we will not assume any cancellation between
$(\mbox{Re}~h_{12}^{(D)})^2$ and $(\mbox{Im}~h_{12}^{(D)})^2$,
so we estimate
\begin{equation}
    |h_{12}^{(D)}| \loe 7\cdot 10^{-8}\;.
\label{5+}
\end{equation}
The contribution of the pseudoscalar sgoldstino exchange
into CP-violating term $m'$
that mixes
$K_1$ and $K_2$ 
(we use the standard notations~\cite{PDG}) 
is also straightforwardly evaluated in chiral theory,
\begin{equation}
  \Delta m' = \mbox{Re}~h_{12}^{(D)}\cdot \mbox{Im}~h_{12}^{(D)}\cdot
  \frac{B_0^2 f_K^2}{m_K(m_K^2 - m_P^2)}\;.
\nonumber
\end{equation}
Requiring that the corresponding contribution into the parameter 
$\epsilon$ of CP-violation in kaon system is smaller than its measured
value (and again neglecting sgoldstino mass), we find
\begin{equation}
   |\mbox{Re}~h_{12}^{(D)}\cdot \mbox{Im}~h_{12}^{(D)}|
    <1.5\cdot10^{-17}\;. 
\label{reim}
\end{equation}

Now, light pseudoscalar sgoldstino can be produced in kaon decays, 
as shown in fig.~2.
\begin{figure}[htb]
\begin{picture}(0,0)%
\epsfig{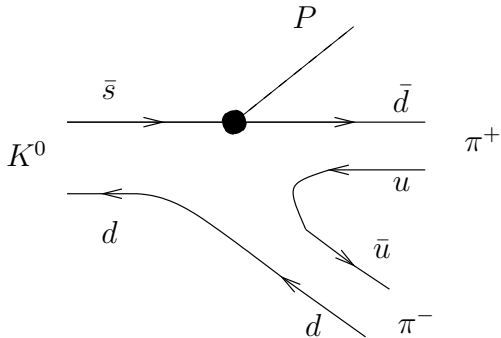}%
\end{picture}%
\setlength{\unitlength}{3947sp}%
\begingroup\makeatletter\ifx\SetFigFont\undefined%
\gdef\SetFigFont#1#2#3#4#5{%
  \reset@font\fontsize{#1}{#2pt}%
  \fontfamily{#3}\fontseries{#4}\fontshape{#5}%
  \selectfont}%
\fi\endgroup%
\begin{picture}(2637,2160)(226,-1786)
\put(226,-631){\makebox(0,0)[lb]{\smash{\SetFigFont{12}{14.4}{\rmdefault}{\mddefault}{\updefault}$K^0$}}}
\put(826,-211){\makebox(0,0)[lb]{\smash{\SetFigFont{12}{14.4}{\rmdefault}{\mddefault}{\updefault}$\bar{s}$}}}
\put(826,-1111){\makebox(0,0)[lb]{\smash{\SetFigFont{12}{14.4}{\rmdefault}{\mddefault}{\updefault}$d$}}}
\put(2106,-1696){\makebox(0,0)[lb]{\smash{\SetFigFont{12}{14.4}{\rmdefault}{\mddefault}{\updefault}$d$}}}
\put(2531,-1206){\makebox(0,0)[lb]{\smash{\SetFigFont{12}{14.4}{\rmdefault}{\mddefault}{\updefault}$\bar{u}$}}}
\put(2656,-786){\makebox(0,0)[lb]{\smash{\SetFigFont{12}{14.4}{\rmdefault}{\mddefault}{\updefault}$u$}}}
\put(2656,-286){\makebox(0,0)[lb]{\smash{\SetFigFont{12}{14.4}{\rmdefault}{\mddefault}{\updefault}$\bar{d}$}}}
\put(2026,239){\makebox(0,0)[lb]{\smash{\SetFigFont{12}{14.4}{\rmdefault}{\mddefault}{\updefault}$P$}}}
\put(3106,-536){\makebox(0,0)[lb]{\smash{\SetFigFont{12}{14.4}{\rmdefault}{\mddefault}{\updefault}$\pi^+$}}}
\put(2686,-1676){\makebox(0,0)[lb]{\smash{\SetFigFont{12}{14.4}{\rmdefault}{\mddefault}{\updefault}$\pi^-$}}}
\end{picture}
\caption{Kaon decay into sgoldstino and pions.}
\label{k-decay}
\end{figure}
Parity conservation implies that these decays 
involve at least two pions in the final state,
\begin{equation}
      K \to \pi \pi P\;.
\nonumber
\end{equation}
The dominant amplitudes come from non-derivative couplings of mesons 
and $P$, so the pions are in $s$-wave state within this approximation. 
Then it is straightforward 
to see that the charged kaon decay $K^{+} \to \pi^{+} \pi^0 P$ is
forbidden at this level\footnote{As the only relevant term in the 
effective Lagrangian is given by Eq.(\ref{4+}), sgoldstino $P$ 
behaves in the process $K\to \pi\pi P$
as a component of an isodoublet. By Bose statistics, $s$-wave state 
of two pions has either isospin $0$ or isospin $2$. In the case of
$K^{+} \to \pi^{+} \pi^0 P$, the isospin-0 state of two pions is 
impossible, so the total isospin of the final state is at least $3/2$.
Hence,  $K^{+} \to \pi^{+} \pi^0 P$ 
is forbidden, in the leading order of derivative expansion,
by the conservation of total isospin.}. 
On the other hand,
the non-derivative couplings of $K^0$, pions and $P$ is straightforward 
to calculate in chiral theory,
\begin{equation}
  {\cal L}_{P,K^0,\pi,\pi} =
  \frac{2B_0}{3f_{\pi}}\cdot P \cdot 
   \left[ h_{12}^{(D)*}\left(K^0 \pi^{+} \pi^{-} + 
    \frac{1}{2} K^{0} \pi^0 \pi^0 \right) +
   h_{12}^{(D)}\left(\bar{K}^0 \pi^{+} \pi^{-} + 
    \frac{1}{2} \bar{K}^{0} \pi^0 \pi^0 \right) \right]\;.
\nonumber
\end{equation}
Neglecting for a moment
the Standard Model CP-violation 
in the neutral kaon system, we find the partial 
widths of $K_L^0$ and $K_S^0$,
\begin{eqnarray}
    \Gamma (K_L^0 \to \pi^{+}\pi^{-} P)
    = 2 \Gamma (K_L^0 \to \pi^0 \pi^0 P)
    = (\mbox{Re}~h_{12}^{(D)})^2 \frac{m_K B_0^2}{576\pi^3 f_{\pi}^2}
\cdot F(m_P, m_{\pi}, m_K)\;,
\nonumber\\
\Gamma (K_S^0 \to \pi^{+}\pi^{-} P)
    = 2 \Gamma (K_S^0 \to \pi^0 \pi^0 P)
    = (\mbox{Im}~h_{12}^{(D)})^2 \frac{m_K B_0^2}{576 \pi^3 f_{\pi}^2}
\cdot F(m_P, m_{\pi}, m_K)\;,
\nonumber
\end{eqnarray}
where $F(m_P, m_{\pi}, m_K)$ is a correction factor
accounting for finite masses
of pions and $P$; at $m_P \approx 0$ it is equal to 
$F\approx0.3$~.

Decays of charged kaons, $K^{+} \to \pi^{+} \pi^0 P$, are due to
isospin violation as well as 
chiral loops and
derivative couplings in the effective meson-sgoldstino Lagrangian. 
These are numerically 
small~\cite{Gasser:1985gg,chiral}, and
the decay amplitudes are somewhat suppressed.
A chiral theory estimate 
(see Appendix 2) 
gives at small $m_P$
\begin{equation}
\mbox{Br} (K^{+} \to \pi^{+} \pi^0 P) \sim 8.5\cdot10^{10}
\cdot |h_{12}^{(D)}|^2\;.
\nonumber
\end{equation}

The ranges of $\mbox{Br}(K\to\pi\pi P)$, allowed by
constraints~(\ref{full}) and (\ref{reim}), depend on the phase of
$h_{12}^{(D)}$. Hence, we have to consider three cases. 

{\it (i)} \underline{Generic phase of $h_{12}^{(D)}$, i.e., 
${\rm Im}~h_{12}^{(D)} \sim {\rm Re}~h_{12}^{(D)}$}. We make use of
the constraint~(\ref{reim}) to obtain the following bounds    
\begin{eqnarray}
\nonumber
{\rm Br}(K_L^0\to\pi^+\pi^-P) &\loe& 2\cdot10^{-3}\;,~~~~
{\rm Br}(K_L^0\to\pi^0\pi^0P)\loe 1\cdot10^{-3}\;,\\
{\rm Br}(K_S^0\to\pi^+\pi^-P) &\loe& 3\cdot10^{-6}\;,~~~~
{\rm Br}(K_S^0\to\pi^0\pi^0P) \loe 1.5\cdot10^{-6}\;,
\label{prediction}
\\
{\rm Br}(K^{+} \to \pi^{+} \pi^0 P) &\loe &1.5\cdot10^{-6}\;.
\nonumber
\end{eqnarray}
We note that in this case the decays of $K_S^0$ are not particularly
interesting, whereas the branching ratio of $K^+\to\pi^+\pi^0 P$ is
about three orders of magnitude lower than ${\rm Br}(K_L^0\to\pi\pi
P)$, 
\begin{equation}
{{\rm Br}(K^+\to\pi^+\pi^0 P)\over{\rm Br}(K_L^0\to\pi^+\pi^- P)}=\half 
{{\rm Br}(K^+\to\pi^+\pi^0 P)\over{\rm Br}(K_L^0\to\pi^0\pi^0 P)}\sim 
10^{-3}\;.
\label{D1*}
\end{equation}

{\it (ii)} \underline{Small phase of $h_{12}^{(D)}$, i.e., 
${\rm Im}~h_{12}^{(D)} \approx 0$}. In this case the
constraint~(\ref{reim}) is irrelevant. The constraint~(\ref{full})
does not imply any meaningful bounds on $\mbox{Br} (K_L^0 \to  \pi \pi
P)$ but gives for $K^+$-decay 
\begin{equation}
 \mbox{Br} (K^{+} \to \pi^{+} \pi^0 P) \loe 4 \cdot10^{-4}\;. 
\label{again}
\end{equation}
The relation between the branching ratios, Eq.~(\ref{D1*}), still
holds. The decays $K_S^0 \to  \pi \pi P$ occur due to CP-violation in
the Standard Model and are suppressed by the square of the SM
parameter $\epsilon$. Bounds on $\mbox{Br} (K_S^0 \to  \pi \pi P)$ are
very strong, 
\begin{equation}
{\rm Br}(K_S^0\to\pi^+\pi^-P)<6\cdot10^{-9}\;,~~~~
{\rm Br}(K_S^0\to\pi^0\pi^0P)<3\cdot10^{-9}\;, 
\nonumber 
\end{equation} 
so in this case search for light pseudoscalar sgoldstinos in
$K_S^0$-decays is hopeless. 

{\it (iii)} \underline{Phase of $h_{12}^{(D)}$ is close to $\pi/2$, i.e., 
${\rm Re}~h_{12}^{(D)} \approx 0$}. Again, the constraint~(\ref{reim})
is irrelevant. In
this case the decays $K_S^0\to\pi\pi P$ are unsuppressed, whereas the
decays $K_L^0\to\pi\pi P$ are suppressed by $\epsilon^2$, as they
originate from the CP-violation in the Standard Model. The
constraint~(\ref{full}) then implies the following bounds
\begin{eqnarray}
\nonumber
{\rm Br}(K_L^0\to\pi^+\pi^-P) &\loe& 3\cdot10^{-6}\;,~~~~
{\rm Br}(K_L^0\to\pi^0\pi^0P)\loe 1.5\cdot10^{-6}\;,\\
{\rm Br}(K_S^0\to\pi^+\pi^-P) &\loe& 1\cdot10^{-3}\;,~~~~
{\rm Br}(K_S^0\to\pi^0\pi^0P) \loe 0.5\cdot10^{-3}\;,
\label{prediction-1} \\
{\rm Br} (K^{+} \to \pi^{+} \pi^0 P) &\loe& 4\cdot10^{-4}\;,
\nonumber
\end{eqnarray}   
Thus, unlike the previous two cases, the search for light pseudoscalar
sgoldstino in three-body decays of $K_S^0$ and $K^+$ is of particular
interest for ${\rm Re}~h_{12}^{(D)} \approx 0$. 

To complete the picture of $K \to \pi \pi P$, let us briefly discuss
the decays of sgoldstinos. Light sgoldstino decays into two photons
or into a pair of charged leptons due to the following terms in
the effective low-energy 
Lagrangian~\cite{bhat,9904367,0001025},
\begin{equation}
   {\cal L}_{P,\gamma} = 
g_{\gamma} \cdot P \cdot \frac{1}{2} \epsilon_{\mu\nu\lambda\rho}
  F^{\mu \nu} F^{\lambda \rho}\, , \,\,\,\,\,\,
   {\cal L}_{P,l} = -h_{ij}^{(L)} \cdot P \cdot 
\bar{l}_i\, i\gamma^5 l_j\;,
\label{Ptogamma}
\end{equation}
where
\begin{equation}
   g_{\gamma} = \frac{1}{2\sqrt{2}}\frac{M_{\gamma \gamma}}{F}\, ,
\nonumber
\end{equation}
$M_{\gamma \gamma}$ is of the order of the photino mass,
and the contribution to 
$h_{ij}^{(L)}$ related to soft slepton masses is
\begin{equation}
 h_{ij}^{(L)} =\frac{1}{\sqrt{2}} 
\frac{\tilde{m}_{L, ij}^{(LR)2}}{F}\;.
\nonumber
\end{equation}
Scalar sgoldstino interacts with photons and leptons in a
similar way.
Almost everywhere in the parameter space, two-photon decays of 
sgoldstinos dominate, although there exist regions where
sgoldstinos decay mostly into $e^{+}e^{-}$-pair~\cite{low}.
Depending on $g_{\gamma}$ and $h_{ij}^{(L)}$, sgoldstino 
decays either
inside or outside the detector: as an example, at 
$M_{\gamma \gamma} \sim 100$ GeV and $\sqrt{F} \sim 1$ TeV,
sgoldstino flies away from the detector if $m_P \loe 10$ MeV;
otherwise it  decays inside the detector
into two photons. At 
$M_{\gamma \gamma} \sim 100$ GeV and $\sqrt{F} \sim 10$ TeV 
the borderline is at $m_P \sim 200$ MeV. Given the uncertainties in 
supersymmetry breaking parameters, no reliable estimate of sgoldstino 
lifetime can be presently made. 

To summarize, very interesting probe of physics of supersymmetry
breaking is the search for processes
\begin{equation}
\begin{array}{rlrl}
K_L^0 \to \pi^{+} \pi^{-}P~&~~~~~~~~~~,&K_L^0 \to \pi^{0} \pi^{0}P~&\\[-8pt] 
\hookrightarrow&\gamma\gamma~~~~~~&\hookrightarrow&\gamma \gamma
\end{array}
\nonumber
\end{equation}
\begin{equation}
    K_L^0 \to \pi^{+} \pi^{-} P \, , \,\,\,
    K_L^0 \to \pi^{0} \pi^{0} P \, , \,\,\,
     \mbox{invisible}\,\, P
\nonumber
\end{equation}
Generally, $K_L^0 - K_S^0$ mass difference does not impose any upper 
bounds on the branching ratios of these decays, but from 
Eq. (\ref{prediction}) we infer that it is most interesting to
search for these decays at the 
level ${\rm Br}(K_L^0\to\pi \pi P) \sim 10^{-3}$ and below.
Search for 
\begin{equation}
\begin{array}{rlrl}
K_L^0 \to \pi^{+} \pi^{-}P~&~~~~~~~~~~,&K_L^0 \to \pi^{0} \pi^{0}P~&\\[-8pt] 
\hookrightarrow&e^{+} e^{-}~~~~~~&\hookrightarrow&e^{+} e^{-}
\end{array}  
\nonumber
\end{equation}
at the same level, 
is also of interest. For most values of the phase of the
sgoldstino-quark coupling $h_{12}^{(D)}$, the branching ratio of the
decay $K^+\to\pi^+\pi P$ is about three orders of magnitude lower than
${\rm Br}(K_L^0\to\pi \pi P)$, whereas the decays  $K_S^0 \to \pi \pi P$
are strongly suppressed. In the special case ${\rm
Re}~h_{12}^{(D)}=0$, 
however, the most promising places to search for light pseudoscalar
sgoldstino are three-body decays of $K^+$ and $K_S^0$. The interesting
ranges of the branching ratios of the decays 
\begin{equation}
    K_S^0 \to \pi \pi P\, , \,\,\,\,\,
 \,\,\,\, (P \to \gamma \gamma,
\mbox{invisible}, e^{+}e^{-})
\nonumber 
\end{equation}
and
\begin{equation}
K^{+} \to \pi^{+} \pi^{0} P
\, , \,\,\,\, (P \to \gamma \gamma,
 \mbox{invisible}, e^{+}e^{-})
\nonumber 
\end{equation} 
start in that case at   
\begin{equation}
{\rm Br}(K^+\to\pi^+\pi^0P) \sim 4\cdot 10^{-4}\;,~~~~
{\rm Br}(K_S^0\to\pi\pi P) \sim 10^{-3}\;,
\nonumber
\end{equation}
while ${\rm Br}(K_L^0 \to \pi \pi P)$ is three orders of magnitude
smaller than ${\rm Br}(K_S^0\to\pi\pi P)$, and two orders of magnitude
smaller than ${\rm Br}(K^+\to\pi^+\pi^0 P)$.

To the best of authors' knowledge, there exists only one experimental 
limit~\cite{newexp}
directly related to the processes under discussion, 
\begin{equation}
{\rm Br}(K^+\to\pi^+\pi^0X(X\to{\rm invisible}))\lesssim 4\cdot
10^{-5}\;,~~~~m_X<80~{\rm MeV}\;;
\label{new-exp}
\end{equation}
for heavier $X$-particle the limit is weaker. This result is important
for the models with light pseudoscalar sgoldstino escaping from the
detector. Indeed, the limit~(\ref{new-exp}) is one order of magnitude
stronger than the bound from kaon mixing in the
cases ({\it ii}) and ({\it
iii}) (see Eqs.~(\ref{again}) and (\ref{prediction-1})). Besides this
direct constraint, there are two other experimental results
which are of interest in our context.
These are measurements of ${\rm
Br}(K_L^0\to\pi^+\pi^-e^+e^-)$=$3.5\cdot10^{-7}$~\cite{kpipiee}
and ${\rm
Br}(K_S^0\to\pi^+\pi^-e^+e^-)$=$5.1\cdot10^{-5}$~\cite{na48}. These
results demonstrate possible sensitivity
of search for sgoldstino decaying
inside a detector into $e^+e^-$ pair, although no limits on the partial
widths of $K_{L,S}^0\to\pi^+\pi^-X(X\to e^+e^-)$ have been obtained
yet.  

~~~3. To get an 
idea of the sensitivity of searches for pseudoscalar sgoldstino
in kaon decays to the fundamental 
parameter of supersymmetric theories, $\sqrt{F}$, let us 
make use of Eq.~(\ref{hDD}). We
recall
that the off-diagonal entries $\tilde{m}^{(LR)2}_{12}$ in the
matrix of squared masses of squarks are constrained irrespectively 
of light sgoldstinos. These constraints again come from FCNC processes and
CP-violation,
but now occuring due to the exchange of superpartners of ordinary 
particles~\cite{masiero}. The bounds depend on 
the masses of squarks, 
$\tilde{m}_Q$, and gluinos, $M_3$. As an example, at
\begin{equation}
    \tilde{m}_Q = M_3 = 500 \, \mbox{GeV}
\label{10*}
\end{equation}
one has~\cite{masiero}
\begin{equation}
     \sqrt{|\mbox{Re}~[(\delta_{12}^{(D)})^2]|} \equiv
    \frac{\sqrt{|(\mbox{Re}~\tilde{m}^{(LR)2}_{D,12})^2-
(\mbox{Im}~\tilde{m}^{(LR)2}_{D,12})^2|}}{m_Q^2} \loe 
     2.7\cdot 10^{-3}\;~~~~{\rm from }~~\Delta m_K\;,
\label{redel12}
\end{equation}
\begin{equation}
     \sqrt{|\mbox{Im}~[(\delta_{12}^{(D)})^2]|} \equiv
    \frac{\sqrt{2|\mbox{Re}~\tilde{m}^{(LR)2}_{D,12}
~\mbox{Im}~\tilde{m}^{(LR)2}_{D,12}|}}{m_Q^2} \loe 
     3.5\cdot 10^{-4}\;~~~~{\rm from }~~\epsilon\;,
\label{imdel12}
\end{equation}
\begin{equation}
     |\mbox{Im}~(\delta_{12}^{(D)})| \equiv
    \frac{|\mbox{Im}~\tilde{m}^{(LR)2}_{D,12}|}{m_Q^2} \loe 
     2.0\cdot 10^{-5}\;~~~~{\rm from }~~\epsilon'/\epsilon\;.
\nonumber
\end{equation}
Note that the limits~(\ref{redel12}), (\ref{imdel12}) 
apply precisely to those  combinations
of ${\rm Re}~\tilde{m}_{D,12}^{(LR)2}$ and 
${\rm Im}~\tilde{m}_{D,12}^{(LR)2}$ which enter Eqs.~(\ref{full}) and 
(\ref{reim}), respectively.

In the case of maximum CP-violation in the squark sector, i.e., at 
$\mbox{Im}~\delta^{(D)}_{12} \sim \mbox{Re}~\delta^{(D)}_{12}$,
one may take, as the best case, $\delta^{(D)}_{12} \sim2\cdot
10^{-5}$; then the constraint
(\ref{reim}) implies
\begin{equation}
    \sqrt{F} > 30~\mbox{TeV}\;.
\nonumber 
\end{equation}
Searches for decays $K_L^0 \to \pi \pi P$ would be sensitive to 
larger $\sqrt{F}$. The values of $\sqrt{F}$ accessible 
to 
these searches are comparable to astrophysical and reactor 
limits~\cite{low} which, however, apply only to models with very 
light sgoldstinos, $m_P \loe 1$~MeV. Also for comparison,
current collider searches for sgoldstinos of masses 
$m_{P,S} \loe 200$~GeV are sensitive to $\sqrt{F}$ at the
level of 1~TeV~\cite{dicus,0001025,Abreu:2000ij}.

If CP is not significantly violated in the squark sector,
$\mbox{Im}~\tilde{m}^{(LR)2}_{12} \approx 0$, then larger
values of flavor violating squark masses are allowed, see
Eq.~(\ref{redel12}). With the above values of squark and gluino
masses, Eq.~(\ref{10*}),
$\delta^{(D)}_{12}$ may be as large as $3\cdot 10^{-3}$.
If this is so,  searches for pseudoscalar sgoldstino
in kaon decays are sensitive to even higher values of the scale of
supersymmetry 
breaking, $\sqrt{F} \goe 85$~TeV. 

We present these estimates for illustration purposes only,
as the allowed range of $\delta^{(D)}_{12}$ depends substantially
on the parameters of superpartners in MSSM\footnote{At 
$M_3=m_Q=1$~TeV one would have 
$\sqrt{F} \goe 120$~TeV and 240~TeV 
instead of $30$~TeV and $85$~TeV, respectively.}.  
It is worth noting that the branching ratios of $K \to \pi\pi P$
decrease rather mildly as $\sqrt{F}$ increases: they scale
as $(\sqrt{F})^{-4}$.

~~~4. Effective interactions (\ref{3*}) lead also to direct mixing of
neutral kaons and pion with the pseudoscalar goldstino,
\begin{eqnarray}
   {\cal L}_{KP-mixing} =  B_0 f_{K} \cdot
    P (h_{12}^{(D)*} K^0 + h_{12}^{(D)} \bar{K}^0)
\label{10a*}\\
   {\cal L}_{\pi P-mixing} = {1\over\sqrt{2}} B_0 f_{\pi} 
    (h_{11}^{(U)} - h_{11}^{(D)})\cdot
    P \pi^0
\label{10a+}
\end{eqnarray}
If sgoldstino is relatively light, these mixing terms 
also give rise to rare kaon decays. Namely,
$K^0-P$-mixing (\ref{10a*})
induces decays
\begin{eqnarray}
K_S^0 \to \gamma \gamma \, , \,\,\,
K_S^0 \to e^{+}e^{-} \, , \,\,\,     
K_S^0 \to \mu^{+} \mu^{-} \, , \,\,\,
\label{sss}\\
K_L^0 \to \gamma \gamma \, , \,\,\,
K_L^0 \to e^{+}e^{-} \, , \,\,\,     
K_L^0 \to \mu^{+} \mu^{-} \, , \,\,\,
\nonumber
\end{eqnarray}
with rates proportional to $(h_{12}^{(D)})^2$ and depending
on couplings of sgoldstinos to photons and leptons,
see Eq.~(\ref{Ptogamma}). 
To illustrate the situation with these decays, let us take
$\sqrt{F}=1$~TeV, $M_{\gamma\gamma}=100$~GeV and 
$m^{(LR)~2}_{L,11} =m_eA_0$, $m^{(LR)~2}_{L,22} =m_\mu A_0$, $A_0=100$~GeV. 
Making use 
of the constraints~(\ref{full}), (\ref{reim}) 
one finds that the branching ratios 
of these decays must be fairly small,
\begin{eqnarray}
  \mbox{Br} (K_{L} \to \gamma \gamma, e^{+}e^{-}, \mu^{+} \mu^{-})
   &< 4.5\cdot10^{-11}\;,~~~~~~~~~
&{\rm Re}~h_{12}^{(D)}\sim {\rm Im}~h_{12}^{(D)}\;,\nonumber\\ 
\mbox{Br} (K_{L} \to \gamma \gamma, e^{+}e^{-}, \mu^{+} \mu^{-})
   &<1.5 \cdot10^{-8}\;,~~~~~~~~~&{\rm Im}~h_{12}^{(D)}\approx 0\;, 
\label{KL}\\\nonumber 
\mbox{Br} (K_{L} \to \gamma \gamma, e^{+}e^{-}, \mu^{+} \mu^{-})
   &< 7.5\cdot10^{-14}\;,~~~~~~~~~&{\rm Re}~h_{12}^{(D)}\approx 0\;.
\end{eqnarray}
\begin{eqnarray}
\mbox{Br} (K_{S} \to \gamma \gamma, e^{+}e^{-}, \mu^{+} \mu^{-})
   &<7.5 \cdot10^{-14}\;,~~~~~~~~~&{\rm Re}~h_{12}^{(D)}\sim {\rm
   Im}~h_{12}^{(D)}\;,
\nonumber\\
\mbox{Br} (K_{S} \to \gamma \gamma, e^{+}e^{-}, \mu^{+} \mu^{-})
  & < 1\cdot10^{-16}\;,~~~~~~~~~&{\rm Im}~h_{12}^{(D)}\approx 0\;,
\nonumber\\\nonumber
\mbox{Br} (K_{S} \to \gamma \gamma, e^{+}e^{-}, \mu^{+} \mu^{-})
   &< 2\cdot10^{-11}\;,~~~~~~~~~&{\rm Re}~h_{12}^{(D)}\approx 0\;. 
\end{eqnarray}	
At larger $\sqrt{F}$ the allowed branching ratios are even
smaller.
The branching ratios of $K_S^0$ decays are below the sensitivity
of current 
experiments.
In other words, available data on the decays (\ref{sss}) 
do not add extra constraints on 
$\mbox{Br} (K_S^0 \to \pi\pi P)$. On the other hand, current
experimental data~\cite{PDG} on
leptonic decays of $K_L^0$
\begin{eqnarray}
\mbox{Br} (K_L^0 \to e^+e^-)=\l9^{+6}_{-4}\r\cdot10^{-12}\;,
\label{KLee}\\
\mbox{Br} (K_{L}^0 \to \mu^{+} \mu^{-})=(7.15\pm0.16)\cdot10^{-9}\;.
\label{KLmumu}
\end{eqnarray}
are comparable and even below than the right hand sides of
Eqs.~(\ref{KL}). This means that Eqs.~(\ref{KLee}) and (\ref{KLmumu})
may give stronger bounds on sgoldstino-quark coupling, as compared to
Eqs.~(\ref{full}) and (\ref{reim}). The bounds coming from
Eqs.~(\ref{KLee}), (\ref{KLmumu}), however,
strongly depend on unknown parameters like $\sqrt{F}$,
$m^{(LR)~2}_{L,ij}$ and sgoldstino masses; in fact, in most part of
the parameter space the analysis of decays $K_L^0 \to e^+e^-$ and $K_{L}^0
\to \mu^{+} \mu^{-}$ either does not lead to new constraints, or leads
to relatively weak bounds on the branching ratios of the three-body
decays of kaons. As an example, with $\sqrt{F}=1$~TeV, 
$m^{(LR)~2}_{L,11} =m_eA_0$,
$A_0=100$~GeV, and assuming ${\rm Br}(P\to e^+e^-)\sim1$, 
one finds from Eq.~(\ref{KLee}) the
following new bound 
\begin{equation}
\mbox{Br} (K_L^0 \to\pi^+\pi^- P(P\to e^+e^-))<3\cdot10^{-4}\;.
\nonumber
\end{equation}
This is to be compared to Eq.~(\ref{prediction}); we see that new bound is 
indeed not much stronger than that coming from Eq.~(\ref{reim}). In more
realistic case of higher $\sqrt{F}$, the bound implied by
Eq.~(\ref{KLee}) is weaker than Eq.~(\ref{prediction}).  

\begin{figure}[htb]
\begin{picture}(0,0)%
\epsfig{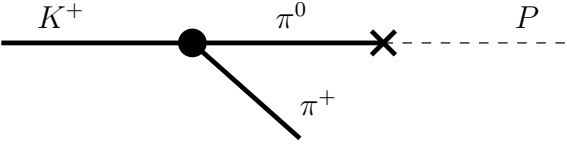}%
\end{picture}%
\setlength{\unitlength}{3947sp}%
\begingroup\makeatletter\ifx\SetFigFont\undefined%
\gdef\SetFigFont#1#2#3#4#5{%
  \reset@font\fontsize{#1}{#2pt}%
  \fontfamily{#3}\fontseries{#4}\fontshape{#5}%
  \selectfont}%
\fi\endgroup%
\begin{picture}(3645,891)(568,-394)
\put(826,335){\makebox(0,0)[lb]{\smash{\SetFigFont{12}{14.4}{\rmdefault}{\mddefault}{\updefault}$K^+$}}}
\put(2476,-211){\makebox(0,0)[lb]{\smash{\SetFigFont{12}{14.4}{\rmdefault}{\mddefault}{\updefault}$\pi^+$}}}
\put(2326,335){\makebox(0,0)[lb]{\smash{\SetFigFont{12}{14.4}{\rmdefault}{\mddefault}{\updefault}$\pi^0$}}}
\put(3826,335){\makebox(0,0)[lb]{\smash{\SetFigFont{12}{14.4}{\rmdefault}{\mddefault}{\updefault}$P$}}}
\end{picture}
\caption{$K^+ \to \pi^+ P$ decay due to $\pi^0$-$P$ mixing.}
\label{charge-kaon-decay}
\end{figure}

The most interesting consequence of $\pi^0 - P$ mixing,
Eq.~(\ref{10a+}), are the decays
\begin{equation}
   K^{+} \to \pi^{+} P\;,~~~~~~K^0 \to \pi^0 P\;,
\label{7*} 
\end{equation}
which involves ordinary weak interaction as shown in 
fig.~\ref{charge-kaon-decay}. 
As the mixing term (\ref{10a+}) does not involve the 
flavor-violating coupling $h_{12}^{(D)}$, the constraints (\ref{5+}),
(\ref{reim})
do not apply here. It is straightforward to see that current 
searches for two-body decays~(\ref{7*}) with $P$
subsequently decaying into $\gamma \gamma$,  $e^{+}e^{-}$, 
$\mu^{+} \mu^{-}$ or flying away from the detector, are sensitive to
the scale of supersymmetry breaking
up to $\sqrt{F} \sim 1\div10$~TeV, depending on squark 
masses, provided that $m_P < (m_K - m_{\pi})$. 

~~~5. For completeness, let us now discuss kaon physics 
in the case when, instead of pseudoscalar sgoldstino $P$,
scalar sgoldstino $S$ is light.
If $m_S < (m_K - m_{\pi})$, both charged and neutral kaons
can decay into $\pi S$, the rates being
\begin{equation}
  \Gamma (K^{+} \to \pi^{+} S) = |h_{12}^{(D)}|^2{B_0^2\over16\pi m_K}
\cdot F'(m_P, m_{\pi}, m_K)\;,
\nonumber 
\end{equation}
\begin{equation}
  \Gamma (K^{0}_L \to \pi^{0} S) = (\mbox{Re}~h_{12}^{(D)})^2
{B_0^2\over16\pi m_K}\cdot F'(m_P, m_{\pi}, m_K)\;,
\nonumber 
\end{equation}
\begin{equation}
  \Gamma (K^{0}_S \to \pi^0 S) = (\mbox{Im}~h_{12}^{(D)})^2
{B_0^2\over16\pi m_K}\cdot F'(m_P, m_{\pi}, m_K)\;. 
\nonumber 
\end{equation}
where $F'$ is again a correction factor; 
$F'\approx 0.9$ at $m_P \approx 0$.

Search for scalars in two-body decays of kaons, $K \to \pi S$,
with $S$ either flying away or decaying into two photons or
lepton pair inside the detector, is well explored area of 
experimental kaon physics. In particular,
depending on the channel of $S$-decay, the existing limits on
$K^{+}$-decay are in the range 
$\mbox{Br} (K^{+} \to \pi^{+}S) <10^{-7}$ to 
$\mbox{Br} (K^{+} \to \pi^{+}S) <10^{-9}$.
These limits are much stronger than the bounds analogous to 
Eqs.~(\ref{5+}) and (\ref{reim}), so the consideration of 
$K_L^0 - K_S^0$ mass difference and CP-violation in kaon system 
is not relevant if $S$ is light.
The sensitivity to $\sqrt{F}$ of searches for rare kaon decays
$K^+ \to \pi^+ S$ is in the range up to $10^3 - 10^4$~TeV,
provided that the scalar sgoldstino mass is smaller than
$(m_K - m_{\pi})$ (see Ref.~\cite{low} for details). Similar analysis
applies to $K_L^0\to\pi^0S$ decay, and searches for rare neutral
kaon decays have sensitivity to $\sqrt{F}$ of the same order.  

Let us again note that if parity is violated substantially in
sgoldstino-quark interactions, the discussion of this section
applies to $P$ as well. The best place to search for {\it both}
$P$ and $S$ (if any of them is lighter than $(m_K - m_{\pi})$)
is then two-body decays $K \to \pi P$, $K \to \pi S$.

~~~6. Finally, as mentioned in Ref.~\cite{Brignole:2000wd}, 
sgoldstinos heavier than
kaons, though cannot be observed in kaon decays, still
contribute to $K_L^0 - K_S^0$ mass difference through
the diagrams of fig.~\ref{k-k-mixing}. If $m_{S,P}$ are
larger than the hadronic scale,
sgoldstino exchange induces effective four-quark interactions
at low energies,
\begin{eqnarray}
{\cal L} &=& \frac{1}{m_S^2}
\left(h_{ij}^{(D)} \bar{d}_i d_j +  h_{ij}^{(U)} \bar{u}_i u_j
\right)
\left(h_{ij}^{(D)*} \bar{d}_i d_j +  h_{ij}^{(U)*} \bar{u}_i u_j
\right) \nonumber \\    
&& +
\frac{1}{m_P^2}
\left(h_{ij}^{(D)} \bar{d}_i\gamma^5 d_j 
+  h_{ij}^{(U)} \bar{u}_i \gamma^5 u_j
\right)
\left(h_{ij}^{(D)*} \bar{d}_i \gamma^5 d_j 
+  h_{ij}^{(U)*} \bar{u}_i \gamma^5 u_j
\right)\;.
\nonumber
\end{eqnarray}
Making use of the Fierz identities
one obtains, within vacuum insertion approximation,
\begin{equation}
\Delta m_K=-\frac{1}{6}\mbox{Re}~(h_{12}^{(D)})^2
\frac{f_K^2 B_0^2}{m_K}
\left(\frac{11}{m_P^2}-
\frac{1}{m_S^2}\right)\;.
\nonumber
\end{equation}
This implies a constraint on $h_{12}^{(D)}$ similar to Eq.~(\ref{5+})
but now depending on $m_S$, $m_P$. Similarly, the mass differences
of neutral $D$- and $B$-mesons constrain $h_{12}^{(U)}$ and
$h_{13}^{(D)}$, respectively. These constraints are summarized in 
Table~\ref{table-mass-mixing} 
\begin{table}[htb]
$$
\begin{array}{|c|c|c|c|}
\hline {\rm Experimental~data}~\cite{PDG} &
m_{P}=10~{\rm GeV}&
m_{P}=100~{\rm GeV}&
m_{P}=1~{\rm TeV}\\
\hline 
\Delta m_K=3.5\cdot10^{-12}~{\rm MeV}
&|\mbox{Re}~(h_{12}^{(D)2})|
< 1\cdot 10^{-12}&|\mbox{Re}~(h_{12}^{(D)})^2|
< 1\cdot 10^{-10}&|\mbox{Re}~(h_{12}^{(D)})^2|
< 1\cdot 10^{-8}\\
&\sqrt{F}>22~{\rm TeV} 
&\sqrt{F}>7~{\rm TeV}&\sqrt{F}>2.2~{\rm TeV}\\
\hline
\Delta m_D<5\cdot10^{-11}~{\rm MeV}
&|\mbox{Re}~(h_{12}^{(U)2})|
< 1.5\cdot 10^{-11}&|\mbox{Re}~(h_{12}^{(U)})^2|
< 1.5\cdot 10^{-9}&|\mbox{Re}~(h_{12}^{(U)})^2|
< 1.5\cdot 10^{-7}\\
&\sqrt{F}>38~{\rm TeV}
&\sqrt{F}>12~{\rm TeV}&\sqrt{F}>3.8~{\rm TeV} \\
\hline
\Delta m_B=3.1\cdot10^{-10}~{\rm MeV}
&|\mbox{Re}~(h_{13}^{(D)2})|
< 5\cdot 10^{-11}&|\mbox{Re}~(h_{13}^{(D)})^2|
< 5\cdot 10^{-9}&|\mbox{Re}~(h_{13}^{(D)})^2|
< 5\cdot 10^{-7}\\
&\sqrt{F}>29~{\rm TeV} 
&\sqrt{F}>9~{\rm TeV} &\sqrt{F}>2.9~{\rm TeV} \\
\hline
\end{array}
$$
\caption{Constraints on SUSY models from measurements of mass
differences of neutral mesons at various $m_P$; flavor violating
terms $\mbox{Re}~(\delta_{12}^{(D)2})
=\mbox{Re}~[(\tilde{m}^{LR 2}_{D_{12}}/\tilde{m}^2_Q)^2]$,
$\mbox{Re}~(\delta_{13}^{(D)2})
=\mbox{Re}~[(\tilde{m}^{LR 2}_{D_{13}}/\tilde{m}^2_Q)^2]$,  
$\mbox{Re}~(\delta_{12}^{(U)2})
=\mbox{Re}~[(\tilde{m}^{LR 2}_{U_{12}}/\tilde{m}^2_Q)^2]$ 
are set equal to their current
limits (see  text) at equal masses of squarks and gluino,
$M_3=\tilde{m}_Q$=500~GeV.}
\label{table-mass-mixing}
\end{table}  
where we assume for definiteness that $m_P < m_S$
(the general case is treated in the same way).

To illustrate the sensitivity of meson mass differences to
$\sqrt{F}$, we again choose parameters according to
Eq.~(\ref{10*}), set 
$\mbox{Re}~(\delta_{ij}^{(U,D)2}) 
\equiv \mbox{Re}~[((m_{U,D;ij}^{(LR)})^2/m_Q^2)^2]$
equal to their experimental limits~\cite{masiero},
namely, $|\mbox{Re}~(\delta_{12}^{(D)2})|^{1/2} = 2.7\cdot 10^{-3}$,
$|\mbox{Re}~(\delta_{12}^{(U)2})|^{1/2} = 3.1\cdot 10^{-2}$,
 $|\mbox{Re}~(\delta_{13}^{(D)2})|^{1/2} = 3.3\cdot 10^{-2}$, 
at $m_Q=M_3=500$~GeV 
and under these assumptions transform the limits on $h_{ij}$
into limits on $\sqrt{F}$. The results are also presented
in Table~\ref{table-mass-mixing}. 

Heavy sgoldstino exchange would  contribute 
also to CP-violation in neutral meson systems.
The corresponding constraints on $\mbox{Im}~(h_{12}^{(D)2})$ and
$\mbox{Im}~(h_{13}^{(D)2})$ are summarized in
Table~\ref{table-CP-mixing}, 
\begin{table}[htb]
$$
\begin{array}{|c|c|c|c|}
\hline {\rm Experimental~data}~\cite{PDG} &
m_{P}=10~{\rm GeV}&
m_{P}=100~{\rm GeV}&
m_{P}=1~{\rm TeV}\\
\hline 
\epsilon_K=2.3\cdot10^{-3}
&|\mbox{Im}~(h_{12}^{(D)})^2|
< 6.5\cdot 10^{-15}&|\mbox{Im}~(h_{12}^{(D)})^2|
< 6.5\cdot 10^{-13}&|\mbox{Im}~(h_{12}^{(D)})^2|
< 6.5\cdot 10^{-11}\\
&\sqrt{F}>28~{\rm TeV} 
&\sqrt{F}>8.7~{\rm TeV}&\sqrt{F}>2.8~{\rm TeV}\\
\hline
\epsilon_B<1\cdot10^{-2}
&|\mbox{Im}~(h_{13}^{(D)})^2|
< 1.4\cdot 10^{-12}&|\mbox{Im}~(h_{13}^{(D)})^2|
< 1.4\cdot 10^{-10}&|\mbox{Im}~(h_{13}^{(D)})^2|
< 1.4\cdot 10^{-8}\\
&\sqrt{F}>85~{\rm TeV} 
&\sqrt{F}>26~{\rm TeV}&\sqrt{F}>8~{\rm TeV}\\
\hline
\end{array}
$$
\caption{Constraints on SUSY models from  CP-violation
in $K^0-\bar{K}^0$ and $B^0-\bar{B}^0$ systems 
at various $m_P$; assumptions entering the bounds on $\sqrt{F}$
are presented in the text.}
\label{table-CP-mixing}
\end{table}  
where we 
again assume
that $P$ is lighter than $S$. The bounds on $\sqrt{F}$
presented in Table~\ref{table-CP-mixing} are  obtained by taking
$m_Q = M_3 = 500$~GeV, setting $\mbox{Im}~(\delta_{12}^{(D)2})$
equal to its upper limit, 
$|\mbox{Im}~(\delta_{12}^{(D)2})|^{1/2}=3.5\cdot 10^{-4}$, and
assuming that
$\mbox{Im}~ \delta_{13}^{(D)}\sim \mbox{Re}~ \delta_{13}^{(D)} \sim 3\cdot
10^{-2}$. We see that the CP-violating parameters in both kaon and
$B$-meson systems are sensitive to the similar range of $\sqrt{F}$
as the mass differences. Notably, sgoldstino exchange may
contribute significantly into CP-violation in $B$-mesons.

Clearly, the above estimates of $\sqrt{F}$ depend on many 
unknown parameters. Still, we conclude that meson mass differences and
parameters of CP-violation are sensitive 
to an interesting range of $\sqrt{F}$ even for relatively heavy 
sgoldstinos.

\vspace{3mm}

The authors are indebted to F. Bezrukov, A. Buras, S. Dubovsky, 
D. Semikoz, L. Stodolsky, V. Zakharov for helpful discussions. This work 
was supported in part under RFBR  grant 99-02-18410, CRDF grant
(award RP1-2103), Swiss Science Foundation grant 7SUPJ062239, and  
by the the Council for
Presidential Grants and State Support of Leading Scientific Schools, grant
00-15-96626.   
\newline

{\bf\Large Appendix 1}
\newline
A prototype model with natural tree-level parity conservation 
in sgoldstino-quark and sgoldstino-gluon interactions
is the supersymmetric
left-right theory whose gauge group is 
$SU(3)_c \times SU(2)_L \times SU(2)_R$. Besides
gauge superfields, it contains goldstino 
superfield (cf.~Ref.~\cite{9904367})
$Z= \l{1\over\sqrt{2}}\l S +iP\r, \psi_z, F_z\r$, 
which is a gauge singlet, quark superfields
$Q$ and $Q^c$ (the generation index is suppressed) transforming as
$(2,1)$ and $(1,2)$ under $SU(2)_L \times SU(2)_R$, two Higgs
bidoublets, both in $(2,2)$ representation, plus other fields
which are required to break left-right symmetry spontaneously
(see Refs.~\cite{otherLR,Mohrev,strongCP} for details).
For  convenience, let us denote the two Higgs bidoublets as 
$\Phi^{(U)}$ and $\Phi^{(D)}$.
The relevant terms in the effective Lagrangian are determined by
a superpotential
\begin{equation}
{\cal W} = FZ + \frac{\sigma}{6} Z^3
+ {\bf Y}^{(i)} Q^T \tau_2 \Phi^{(i)} \tau_2 Q^c
+  {\bf y}^{(i)}Z Q^T \tau_2 \Phi^{(i)} \tau_2 Q^c
+ \dots \, , \,\,~~~~~ i=U,D \,\, ,
\label{A11*}
\end{equation}
the gauge kinetic functions 
\begin{equation}
    f_3 = \frac{1}{g_3^2} (1 + 2\eta_3 Z + \dots)\,\, ,
\label{A11+}
\end{equation}
\begin{equation}
    f_{L,R} = \frac{1}{g_{L,R}^2} (1 + 2\eta_{L,R} Z + \dots) \,\, ,
\end{equation}
and the K\"ahler potential
\begin{equation}
   K = K_{can} + K_{non-ren.}\;,
\nonumber
\end{equation}
the latter containing
 canonical kinetic terms for chiral superfields and also
non-renormalizable terms,
\begin{equation}
  K_{non-ren.} = - \alpha_Z \frac{|Z|^4}{4} - 
{\bf B}_Q |Z|^2 Q Q^{*} - {\bf B}_{Q^c} |Z|^2 Q^c Q^{c*}
 + \dots 
\nonumber
\end{equation}
The supersymmetric Lagrangian is then
\begin{equation}
  {\cal L} = \int~d^2\theta~({1\over 4}f_a W_a W_a + {\cal W})
  + \int~d^2\bar{\theta}~({1\over 4}f_a^{*} W_a^{*} W_a^{*} + {\cal W}^{*})
  + \int~d^2\theta~d^2\bar{\theta}~K\;.
\label{A12*}
\end{equation}
Left-right symmetry is realized as an involution, with the interchange of
$SU(2)_L \leftrightarrow SU(2)_R$,
\begin{eqnarray}
   \theta &\leftrightarrow& \bar{\theta}\nonumber\\
  Q &\leftrightarrow& Q^{c *}\nonumber\\
   \Phi^{(i)} &\leftrightarrow& \Phi^{(i)\dagger}\nonumber\\
   Z& \leftrightarrow& Z^{*}\nonumber\\
   W_L &\leftrightarrow& W_R^{*}\nonumber\\\nonumber
   W_{SU(3)} &\leftrightarrow& W_{SU(3)}^{*}
\nonumber
\end{eqnarray}
This symmetry imposes the following constraints on the parameters 
of the model,
\begin{eqnarray}
   F &=& F^{*} \nonumber \\
   \sigma &=& \sigma^{*} \nonumber \\
    {\bf Y}^{(i)} &=& {\bf Y}^{(i) \dagger} \nonumber \\
      {\bf A}^{(i)} &=& {\bf A}^{(i) \dagger} \,\, , \,\, ~~~~~
     i = U,D   
\label{A13+} \\
    \eta_3 &=& \eta_3^{*} \nonumber \\
    \eta_{L,R} &=& \eta_{R,L}^{*} \nonumber \\
    {\bf B}_Q &=& {\bf B}_{Q^c} \nonumber
\end{eqnarray}
where
\[
 {\bf A}^{(i)} \equiv F{\bf y}^{(i)}\;.
\] 
To solve the strong CP-problem, one also requires
\begin{equation}
  \eta_L = \eta_L^{*} \, , \,\,\,\,\, \,\, \eta_R = \eta_R^{*}\;.
\label{A13*}
\end{equation}
All 
these conditions are valid above the scale $M_R$ at which the
left-right symmetry is spontaneously broken. Below this scale,
the model reduces to MSSM with the sgoldstino superfield and with
relations (\ref{A13+}) valid at the tree level.

There exists a local supersymmetry breaking minimum of the scalar
potential at which the $F$-component of the sgoldstino 
superfield has non-zero value,
\begin{equation}
    \la F_z\ra = F\;,
\nonumber
\end{equation}
Due to supersymmetry breaking, scalar and pseudoscalar components of $Z$
acquire masses,
\begin{equation}
    m_S^2 = \alpha_z F^2 + \sigma F\,\, , \,\,\,\,\,\,\,\,
    m_P^2 = \alpha_z F^2 - \sigma F\;. 
\nonumber
\end{equation}
Soft masses of squarks and gauginos, as well as trilinear soft
terms are also generated. In particular, the gluino mass is
\begin{equation}
   M_{\lambda_3} = \eta_3 F\;,
\label{A14+}
\end{equation}
and trilinear soft terms involve
\begin{equation}
   {\cal L}_{soft} = {\bf A}^{(i)} \tilde{Q}^T \tau_2 \Phi^{(i)}
        \tau_2 \tilde{Q}^c \,\, , \,\,~~~~~ i = U,D
\label{A14*}
\end{equation}
where $\tilde{Q}$ and $\tilde{Q}^c$ are squark fields.

Finally, breaking of $SU(2)_L$ is arranged in such a way that 
$\Phi^{(U,D)}$ obtain real vacuum expectation values~\cite{strongCP},
\begin{equation}
    \Phi^{(U)} = \l\begin{array}{cc}0&0\\0&\upsilon_U\end{array}\r\;,~~~ 
    \Phi^{(D)} =
    \l\begin{array}{cc}\upsilon_D&0\\0&0\end{array}\r\;,
~~~~~~~v_{U,D} = \mbox{real}\;.
\nonumber
\end{equation}
The soft term (\ref{A14*}) then produces left-right entries in
the matrix of squared masses of squarks,
\begin{equation}
    {\bf \tilde{m}}_{D}^{(LR)2} = {\bf A}^{(D)} v_D\,\, , \,\,\,\,\,\,\,
{\bf \tilde{m}}_{U}^{(LR)2} = {\bf A}^{(U)} v_U\;.
\label{A14**}
\end{equation}

The interactions of sgoldstinos $S$ and $P$ with ordinary quarks
and gluons are read off from Eq.~(\ref{A12*}). Sgoldstino-gluon
interaction is due to the second term in Eq.~(\ref{A11+}),
\begin{equation}
    {\cal L}_{S,P,G} = 
-\frac{1}{2\sqrt{2}} \eta_3 S G_{\mu\nu} G^{\mu\nu}
+ \frac{1}{4\sqrt{2}} \eta_3 P \epsilon_{\mu\nu\lambda\rho}
G^{\mu\nu} G^{\lambda\rho}\;.
\label{A15*}
\end{equation}
Equations (\ref{A14+}) and (\ref{A15*}) illustrate the relations between 
the gaugino mass and sgoldstino couplings to gauge bosons, cf.
Eq.~(\ref{Ptogamma}). Note that the relations (\ref{A13+}) indeed ensure
parity conservation at the tree level.

The couplings of sgoldstinos to quarks come from the last term in 
Eq.~(\ref{A11*}). In view of Eq.~(\ref{A14**}) they can be written
in the following way,
\begin{eqnarray}
  {\cal L}_{S,q} &=& -\l h_{ij}^{(D)} S \bar{d}_i d_j +
\tilde{h}_{ij}^{(D)} S \bar{d}_i\, i \gamma^5 d_j \,\, + (d \to u)\r\;,
\nonumber \\
 {\cal L}_{P,q} &=& -\l h_{ij}^{(D)} P \bar{d}_i \, i \gamma^5 d_j +
\tilde{h}_{ij}^{(D)} S \bar{d}_i d_j \,\, + (d \to u)\r\;,
\nonumber
\end{eqnarray}
where we restored the generation indices and denoted
\begin{eqnarray}
 h_{ij}^{(D)} &=& \frac{1}{2\sqrt{2}}
 \frac{\tilde{m}_{D,ij}^{(LR)2} +[\tilde{m}_{D}^{(LR)2}]^{\dagger}_{ij}}{F}
 \equiv \frac{1}{\sqrt{2}} \frac{v_D}{F} 
\frac{A^{(D)}_{ij} + [A^{(D)\dagger}]_{ij}}{2}\;, \nonumber \\
\tilde{h}_{ij}^{(D)} &=& \frac{1}{2\sqrt{2}}
 \frac{\tilde{m}_{D,ij}^{(LR)2} -[\tilde{m}_{D}^{(LR)2}]^{\dagger}_{ij}}{iF}
 \equiv \frac{1}{\sqrt{2}} \frac{v_D}{F} 
\frac{A^{(D)}_{ij} - [A^{(D)\dagger}]_{ij}}{2i}\;.
\nonumber
\end{eqnarray}
Due to Eq.~(\ref{A13+}), the matrices $\tilde{m}_{U, ij}^{(LR)2}$,
$\tilde{m}_{D, ij}^{(LR)2}$ 
are Hermitian above the scale $m_R$, and we come to
Eqs.~(\ref{3*}), (\ref{hDD}), (\ref{3**}). Hence, parity is conserved 
in sgoldstino-quark interactions at the tree level.

Below the scale $m_R$ of left-right symmetry breaking, relations 
(\ref{A13+}) are no longer valid. In particular, ${\bf A}^{(U,D)}$
receive non-Hermitian contributions in loops, and parity-violating couplings 
are generated.  At the one-loop level, contributions to the non-Hermitian
parts of ${\bf A}^{(U,D)}$ come from diagrams involving (s)quarks
and Higgs(ino)s (assuming the validity\footnote{If Eq.~(\ref{A13*}) does not 
hold at the tree level,  there are also one-loop
contributions involving gauginos. These contributions, 
however, are proportional to
quark Yukawa couplings, $\Delta {\bf A}^{(D)} \propto {\bf Y}^{(D)}$,
and hence are diagonal in the basis where ${\bf Y}^{(D)}$ is diagonal.
As the kaon decays occur due to non-zero $A^{(D)}_{12}$, the latter
contributions are irrelevant for our purposes.}
of Eq.~(\ref{A13*})).
One obtains for the non-Hermitian part of ${\bf A}^{(D)}$
\begin{equation}
\frac{{\bf A}^{(D)} -{\bf A}^{(D)\dagger}}{2}
  =\frac{1}{32\pi^2} ({\bf Y}^{(U)} {\bf Y}^{(U)\dagger}{\bf A}^{(D)}
  + {\bf A}^{(U)} {\bf Y}^{(U)\dagger}{\bf Y}^{(D)} - \, \mbox{h.~c.})
\cdot \log{\frac{M_R^2}{M_{weak}^2}}\;,  
\nonumber
\end{equation}
where $M_{weak}$ is the electroweak scale.
The tree-level values of ${\bf Y}^{(U,D)}$ and ${\bf A}^{(U,D)}$
entering the right hand side are Hermitian; by unitary rotations of
quarks one  makes ${\bf Y}^{(D)}$ diagonal,  
${\bf Y}^{(D)}={\bf Y}^{(D)}_{diag}$, whereas  ${\bf Y}^{(U)}$
takes the form    
\begin{equation}
  {\bf Y}^{(U)} =  {\bf V} {\bf Y}^{(U)}_{diag}\;,
\nonumber
\end{equation}
where ${\bf V}$ is the CKM matrix.

Parity-violating coupling relevant for kaon decays is proportional to
$({\bf A}^{(D)} - {\bf A}^{(D)\dagger})_{12}$. The contribution to 
this term that does not contain small Yukawa couplings of
$d$-, $u$-, $s$- and $c$-quarks is
\begin{equation}
   \frac{1}{2} ({\bf A}^{(D)} - {\bf A}^{(D)\dagger})_{12} =
   \frac{Y_t^2}{32\pi^2} \l V_{13}V^{*}_{i3}A^{(D)}_{i2} -
                         A^{(D)*}_{1i}V_{i3}V^{*}_{32}\r
\cdot \ln{\frac{M_R^2}{M_{weak}^2}}\;.
\nonumber
\end{equation}
Assuming that $A^{(D)}_{12}$ is not particularly small  compared to
$A^{(D)}_{11}$ and $A^{(D)}_{22}$, we find that the largest 
contributions here are
\begin{equation}
 \frac{\tilde{h}^{(D)}_{12}}{h_{12}^{(D)}} \equiv 
\frac{({\bf A}^{(D)} - {\bf A}^{(D)\dagger})_{12}}{2A_{12}^{(D)}}
   = \frac{Y_t^2}{32\pi^2} \left( V_{cb}V_{ts} \e^{i\phi_1}
 + \frac{|A^{(D)}_{13}|}{|A^{(D)}_{12}|} V_{tb} V_{ts} \e^{i\phi_2}\right)
  \cdot \ln{\frac{M_R^2}{M_{weak}^{2}}}\;,
\nonumber
\end{equation}
where the phases $\phi_{1,2}$ are irrelevant for our estimates.
With $Y_t\sim 1$ and $|A^{(D)}_{13}| \sim |A^{(D)}_{12}|$ we obtain
\begin{equation}
    \frac{|\tilde{h}^{(D)}_{12}|}{|h_{12}^{(D)}|} \sim 10^{-3}\;.
\nonumber
\end{equation}
Assuming, by analogy to quark mixing, mild hierarchy
$|A^{(D)}_{13}| / |A^{(D)}_{12}| \loe (\mbox{a~few~})\cdot10^{-2}$ 
we would find instead
\begin{equation}
   \frac{|\tilde{h}^{(D)}_{12}|}{|h_{12}^{(D)}|} 
\loe (\mbox{a~few~})\cdot10^{-5}\;.
\label{A1fin}
\end{equation}
Similar contributions to $\tilde{h}^{(D)}_{12}$
are expected from threshold effects.
In terms of the branching ratios of kaon decays, this 
means 
\begin{eqnarray}
\frac{\mbox{Br}(K_L^0 \to \pi P)}{\mbox{Br}(K_L \to \pi \pi P)} 
&\loe& 10^{-7} \,\, ,\nonumber \\
\frac{\mbox{Br}(K_S^0 \to \pi P)}{\mbox{Br}(K_S \to \pi \pi P)} 
&\loe& 10^{-7}\;, ~~~~~~ 
\mbox{Im}~h_{12}^{(D)} \sim  \mbox{Re}~h_{12}^{(D)}\;,
\nonumber \\
\frac{\mbox{Br}(K^+ \to \pi^+ P)}{\mbox{Br}(K^+ \to \pi^+ \pi^0 P)} 
&\loe& 3\cdot 10^{-5}\;,\nonumber 
\end{eqnarray}
Of course, these estimates depend on many unknown parameters,
but they do suggest that the two-body decays of kaons 
are suppressed as compared to three-body decays.

Yet another source of parity violation in sgoldstino-quark 
interactions is the radiatively induced non-Hermitian part in
the
Yukawa matrix ${\bf Y}^{(D)}$. Its effect, however, can be shown
to be even smaller than the estimate (\ref{A1fin}), cf. 
Ref.~\cite{pospelov}.
\newline

{\bf\Large Appendix 2}
\newline
Charged kaon decay into pseudoscalar sgoldstino and pions is
suppressed in chiral theory. There are two types of contributions
into this process. The first one is due to isospin violation and 
originates from tree-level diagram
proportional to the mass difference between up- and down-quarks,
\begin{equation}
{\cal L}={B_0\l m_u-m_d\r\over 2\sqrt{2}f_\pi^2}K^+\bar{K}^0\pi^-\pi^0\;.
\label{12*}
\end{equation}
This leads to $K^+\to\pi^-\pi^0P$ decay via $\bar{K}^0-P$ mixing, 
$$
{\cal L}={B_0^2f_Kh_{12}^{(D)}\l m_u-m_d\r\over 
2\sqrt{2}f_\pi^2m_K^2}K^+\pi^-\pi^0P
$$
$$
\Gamma(K^+\to\pi^-\pi^0P)={B_0^4f_K^2(m_u-m_d)^2\over 
8^4\pi^3f_\pi^4m_K^4}|h_{12}^D|^2F(m_P,m_\pi,m_K)
$$
where $F(0,0,m_K)=1$, $F(0,m_\pi,m_K)=0.3$. For $m_d-m_u=5$~MeV we
obtain 
\begin{equation}
{\rm Br}(K^+\to\pi^-\pi^0P)=1\cdot10^{10}|h_{12}^{(D)}|^2\;.
\label{isospin-violation}
\end{equation}

Similar contribution from kinetic term 
\begin{eqnarray}
\nonumber
{\cal L}={1\over\sqrt{2}f_\pi^2}\Biggl(\l\pi^-\d_\mu\pi^0-\pi^0\d_\mu\pi^-\r
\l K^+\d^\mu\bar{K}^0-\bar{K}^0\d^\mu K^+\r\\+{\sqrt{3}\over4}\l
\l\pi^0\d_\mu K^+-K^+\d_\mu\pi^0\r\l\bar{K}^0\d^\mu\pi^--\pi^-\d^\mu
\bar{K}^0\r+\l\pi^-\d_\mu K^+-K^+\d_\mu\pi^-\r
\l\bar{K}^0\d^\mu\pi^0-\pi^0\d^\mu
\bar{K}^0\r\r\Biggr)
\nonumber
\end{eqnarray}
is suppressed at least by an order of magnitude as compared to 
Eq.~(\ref{isospin-violation}). 

The second type of contributions occurs in 
the next-to-leading order in momenta. 
Generically, the next-to-leading amplitudes are sums of
chiral loops and tree level contributions due to explicit
higher order terms in the chiral Lagrangian~\cite{Gasser:1985gg},
\begin{equation}
{\cal L}_{int}=L_5~{\rm Tr}\left[\d_\mu U^{\dagger}\d_\mu U\l
iU^{\dagger}{\bf h}^{(D)}-i{\bf h}^{(D)\dagger}U\r\right] \cdot 2B_0P\;,
\nonumber
\end{equation}
where $3\times 3$ matrix $U$ describes light mesons.
The dimensionless coupling constant
$L_5$ depends on the normalization point $\mu$; it runs according to
\[
   L_5(\mu_2) = L_5(\mu_1) +
     \frac{3}{128 \pi^2} \log{\frac{\mu_1}{\mu_2}}
\]
and its value at the $\rho$-meson mass is
$L_5(m_{\rho})=(1.4 \pm 0.5) \cdot10^{-3}$~\cite{chiral}.
The dependence of $L_5$ on $\mu$ is cancelled out in
physical quantities by chiral loops~\cite{Gasser:1985gg}.   

The part of ${\cal L}_{int}$ relevant for the decay $K^+\to\pi^+\pi^0 P$
is
\begin{equation}
{\cal L}_{K^+,P,\pi^0,\pi^-}=L_5~{8\sqrt{2}B_0\over
f_\pi^3}h_{12}^{(D)}P\d_\mu K^+\l\pi^-\d_\mu\pi^0-\pi^0\d_\mu\pi^-\r\;.
\label{charge-eff-2}
\end{equation}
The calculation of chiral loops is performed by making use of general
technique developed in Ref.~\cite{Gasser:1985gg}. In standard
notations~\cite{Gasser:1985gg} the amplitude reads  
\begin{eqnarray}
A={B_0h_{12}^{(D)}\over64\sqrt{2}\pi^2f_\pi^3}\Biggl(
\bar{J}\l
K,\eta,(p_P+p_{\pi^+})^2\r\l m_KE_{\pi^0}
+{1\over2}{\l m_K^2-m_\pi^2\r\l m_\eta^2-m_K^2\r\over\l
p_P+p_\pi^+\r^2}
\r\nonumber\\
-\bar{J}\l
K,\eta,(p_P+p_{\pi^0})^2\r\l m_KE_{\pi^+}
+{1\over2}{\l m_K^2-m_\pi^2\r\l m_\eta^2-m_K^2\r\over\l
p_P+p_\pi^0\r^2}
\r\nonumber\\
+\bar{J}\l 
K,\pi,(p_P+p_{\pi^+})^2\r\l5m_KE_{\pi^0}
+{3\over2}{\l m_K^2-m_\pi^2\r^2\over\l p_P+p_\pi^+\r^2}
\r
-\bar{J}\l
K,\pi,(p_P+p_{\pi^0})^2\r\l5m_KE_{\pi^+}
+{3\over2}{\l m_K^2-m_\pi^2\r^2\over\l p_P+p_\pi^0\r^2}
\r
\nonumber\\\nonumber
+2\l E_{\pi^+}-E_{\pi^0}\r\l5k(K,\pi,\mu)+k(K,\eta,\mu)-512\pi^2L_5(\mu)\r
\Biggr)
\end{eqnarray}
where 
\begin{eqnarray}
\bar{J}\l A,B,s\r=2+\l{\Delta\over
s}-{\Sigma\over\Delta}\r\ln{m_A^2\over m_B^2}-{\nu\over
s}\ln{\l s+\nu\r^2-\Delta^2\over\l s-\nu\r^2-\Delta^2}\;,\nonumber\\ 
\Sigma=m_A^2+m_B^2\;,~~~\Delta=m_B^2-m_A^2\;,\nonumber~~~~~
\nu^2=s^2+m_A^4+m_B^4-2s\l m_A^2+m_B^2\r-2m_A^2m_B^2\;,\\\nonumber
k(A,B,\mu)={m_A^2\ln{m_A^2\over \mu^2}-m_B^2\ln{m_B^2\over\mu^2}\over
m_A^2-m_B^2}\;.
\end{eqnarray}
and $E_A$ is energy of particle $A$ in the rest frame of decaying 
kaon. The dependence on the normalization point $\mu$ indeed cancels
out. In the limit of massless $P$ one obtains 
\begin{equation}
{\rm Br}(K^+\to\pi^-\pi^0P)=7.5\cdot10^{10}|h_{12}^{(D)}|^2
\label{one-loop}
\end{equation}

We note further that the isospin violation,
Eq.~(\ref{12*}), gives rise to $s$-wave amplitude,
whereas the chiral loops together with the term~(\ref{charge-eff-2})
lead to the $p$-wave amplitude. Thus, these two amplitudes do not
interfere, and we merely sum up the
contributions~(\ref{isospin-violation}) and (\ref{one-loop}). We
obtain finally 
\begin{equation}
{\rm Br}(K^+\to\pi^-\pi^0P)=8.5\cdot10^{10}|h_{12}^{(D)}|^2\;.
\nonumber
\end{equation}
We recall that this result is valid at small mass of the pseudoscalar
sgoldstino $P$. 

\def\ijmp#1#2#3{{\it Int. Jour. Mod. Phys. }{\bf #1~} (19#2) #3}
\def\pl#1#2#3{{\it Phys. Lett. }{\bf B#1~} (19#2) #3}
\def\zp#1#2#3{{\it Z. Phys. }{\bf C#1~} (19#2) #3}
\def\prl#1#2#3{{\it Phys. Rev. Lett. }{\bf #1~} (19#2) #3}
\def\rmp#1#2#3{{\it Rev. Mod. Phys. }{\bf #1~} (19#2) #3}
\def\prep#1#2#3{{\it Phys. Rep. }{\bf #1~} (19#2) #3}
\def\pr#1#2#3{{\it Phys. Rev. }{\bf D#1~} (19#2) #3}
\def\np#1#2#3{{\it Nucl. Phys. }{\bf B#1~} (19#2) #3}
\def\mpl#1#2#3{{\it Mod. Phys. Lett. }{\bf A#1~} (19#2) #3}
\def\arnps#1#2#3{{\it Annu. Rev. Nucl. Part. Sci. }{\bf #1~} (19#2) #3}
\def\sjnp#1#2#3{{\it Sov. J. Nucl. Phys. }{\bf #1~} (19#2) #3}
\def\jetp#1#2#3{{\it JETP Lett. }{\bf #1~} (19#2) #3}
\def\app#1#2#3{{\it Acta Phys. Polon. }{\bf #1~} (19#2) #3}
\def\rnc#1#2#3{{\it Riv. Nuovo Cim. }{\bf #1~} (19#2) #3}
\def\ap#1#2#3{{\it Ann. Phys. }{\bf #1~} (19#2) #3}
\def\ptp#1#2#3{{\it Prog. Theor. Phys. }{\bf #1~} (19#2) #3}
\def\spu#1#2#3{{\it Sov. Phys. Usp.}{\bf #1~} (19#2) #3}
\def\apj#1#2#3{{\it Ap. J.}{\bf #1~} (19#2) #3}
\def\epj#1#2#3{{\it Eur.\ Phys.\ J. }{\bf C#1~} (19#2) #3}
\def\pu#1#2#3{{\it Phys.-Usp. }{\bf #1~} (19#2) #3}
\def\nc#1#2#3{{\it Nuovo Cim. }{\bf A#1~} (19#2) #3}

\end{document}